\documentclass[]{spie}  

\usepackage{amsmath,amsfonts,amssymb}
\usepackage{graphicx}
\usepackage[colorlinks=true, allcolors=blue]{hyperref}
\usepackage{soul}

\graphicspath{{figures/}}

\title{MagAO-X: current status and plans for Phase II}

\author[a]{Jared R. Males}
\author[a]{Laird M. Close}
\author[a]{Sebastiaan Haffert}
\author[a]{Joseph D. Long}
\author[a,b]{Alexander D. Hedglen}
\author[a]{Logan Pearce}
\author[e]{Alycia J. Weinberger}
\author[a,b,c,d]{Olivier Guyon}
\author[a,b]{Justin M. Knight}
\author[a,b]{Avalon McLeod}
\author[a,b]{Maggie Kautz}
\author[a,b]{Kyle Van Gorkom}
\author[a,b]{Jennifer Lumbres}
\author[a,b]{Lauren Schatz}
\author[a,b]{Alex Rodack}
\author[a,b]{Victor Gasho}
\author[a,b]{Jay Kueny}
\author[a,b]{Warren Foster}

\affil[a]{Steward Observatory, University of Arizona} 
\affil[b]{James C. Wyant College of Optical Sciences, University of Arizona}
\affil[c]{Subaru Telescope, National Astronomical Observatory of Japan}
\affil[d]{Astrobiology Center, National Institutes of Natural Sciences, Japan}
\affil[e]{Earth and Planets Laboratory, Carnegie Institution for Science}

\authorinfo{Further author information: (Send correspondence to J.R.M.)\\J.R.M.: E-mail: jrmales@arizona.edu}

\begin{document} 
\maketitle

\begin{abstract}
   We present a status update for MagAO-X, a 2000 actuator, 3.6 kHz adaptive optics and coronagraph system for the Magellan Clay 6.5 m telescope.  MagAO-X is optimized for high contrast imaging at visible wavelengths.  Our primary science goals are detection and characterization of Solar System-like exoplanets, ranging from very young, still-accreting planets detected at H-alpha, to older temperate planets which will be characterized using reflected starlight.  First light was in Dec, 2019, but subsequent commissioning runs were canceled due to COVID-19.  In the interim, MagAO-X has served as a lab testbed.  Highlights include implementation of several focal plane and low-order wavefront sensing algorithms, development of a new predictive control algorithm, and the addition of an IFU module.  MagAO-X also serves as the AO system for the Giant Magellan Telescope High Contrast Adaptive Optics Testbed.  We will provide an overview of these projects, and report the results of our commissioning and science run in April, 2022. Finally, we will present the status of a comprehensive upgrade to MagAO-X to enable extreme-contrast characterization of exoplanets in reflected light.  These upgrades include a new post-AO 1000-actuator deformable mirror inside the coronagraph, latest generation sCMOS detectors for wavefront sensing, optimized PIAACMC coronagraphs, and computing system upgrades.  When these Phase II upgrades are complete we plan to conduct a survey of nearby exoplanets in reflected light.

\end{abstract}


\section{INTRODUCTION}
\label{sec:intro}  

MagAO-X\cite{2018SPIE10703E..09M,2020SPIE11448E..4LM} is an ``extreme'' adaptive optics (ExAO) system on the Magellan Clay 6.5 m telescope at Las Campanas Observatory (LCO), in Chile.  It is optimized for high-contrast science at visible-to-near-IR wavelengths.  MagAO-X is also employed as a testbed for adaptive optics (AO) and segment phasing technologies for the Giant Magellan Telescope (GMT), and as a pathfinder for an ExAO instrument planned for GMT called GMagAO-X.

We have successfully completed the first two of several commissioning runs.  First-light was in Dec, 2019.  Runs in 2020 and 2021 were postponed due to the pandemic.  The second commissioning run, as well as initial science operations, occurred in April, 2022.  The main science goal of this initial phase is a search for and characterization of young accreting planets in H$\alpha$ orbiting nearby T Tauri and Herbig Ae/Be stars \cite{2020AJ....160..221C}.   This project uses the simultaneous differential imaging (SDI) mode of MagAO-X to study where and how this  population of low-mass outer extrasolar giant planets (EGPs) form.  Additional near-term science cases for MagAO-X include: circumstellar disk characterization; young EGP characterization in the red-optical/near-IR; spatially resolved stellar surface imaging at high spectral resolution; characterization of tight binary star systems; \textit{Kepler} and \textit{TESS} followup; and high spatial resolution imaging of asteroid surfaces and asteroid companion searches. 

The ultimate goal for MagAO-X is the characterization of nearby temperate exoplanets in reflected light.  Achieving the most demanding wavefront control precision required to perform such observations will require continued improvements and upgrades as MagAO-X commissioning proceeds.  With a 2040 actuator high order deformable mirror (DM) being controlled at up to 3.6 kHz by a pyramid wavefront sensor (PWFS), MagAO-X is optimized for very high Strehl at short wavelengths.  This excellent wavefront quality enables narrow-angle high contrast imaging with coronagraphs.  Wavefront quality is further augmented using coronagraphic low-order and non-common path wavefront sensing and control (WFS\&C).

In what follows we present a brief overview of the instrument, including the design, specifications, and concept of operations.  Following this we report results from our successful second commissioning period, as well as its ongoing use as a laboratory testbed\cite{2020SPIE11448E..2XH, 10.1117/1.JATIS.8.2.021513, 10.1117/1.JATIS.8.2.021515, Demers_2022,Hedglen_2022,Kautz_2022}. Finally we discuss future plans for MagAO-X and in-progress upgrades.

\section{INSTRUMENT OVERVIEW}
\label{sec:design}
Here we present a brief overview of the design and specifications of MagAO-X.  See our previous SPIE contributions for more detailed treatments \cite{2018SPIE10703E..09M,2018SPIE10703E..5AV,2018SPIE10703E..55H,2018SPIE10703E..4ZL,2018SPIE10703E..4YC,2018SPIE10703E..2QK,2018SPIE10703E..2NR,2018SPIE10703E..21S,2018SPIE10703E..1TM,2018SPIE10706E..5OK,2018SPIE10703E..1EG,2020SPIE11448E..4LM,2020SPIE11448E..0UC}  In addition, the complete preliminary design review (PDR) documentation is available at \url{https://magao-x.org/docs/handbook/appendices/pdr/}, and results of laboratory integration and preparation for shipment can be bound in the pre-ship review (PSR) documentation:  \url{https://magao-x.org/docs/handbook/appendices/psr/index.html}.

\begin{figure}[h]
   \centering
   \includegraphics[width=6.5in]{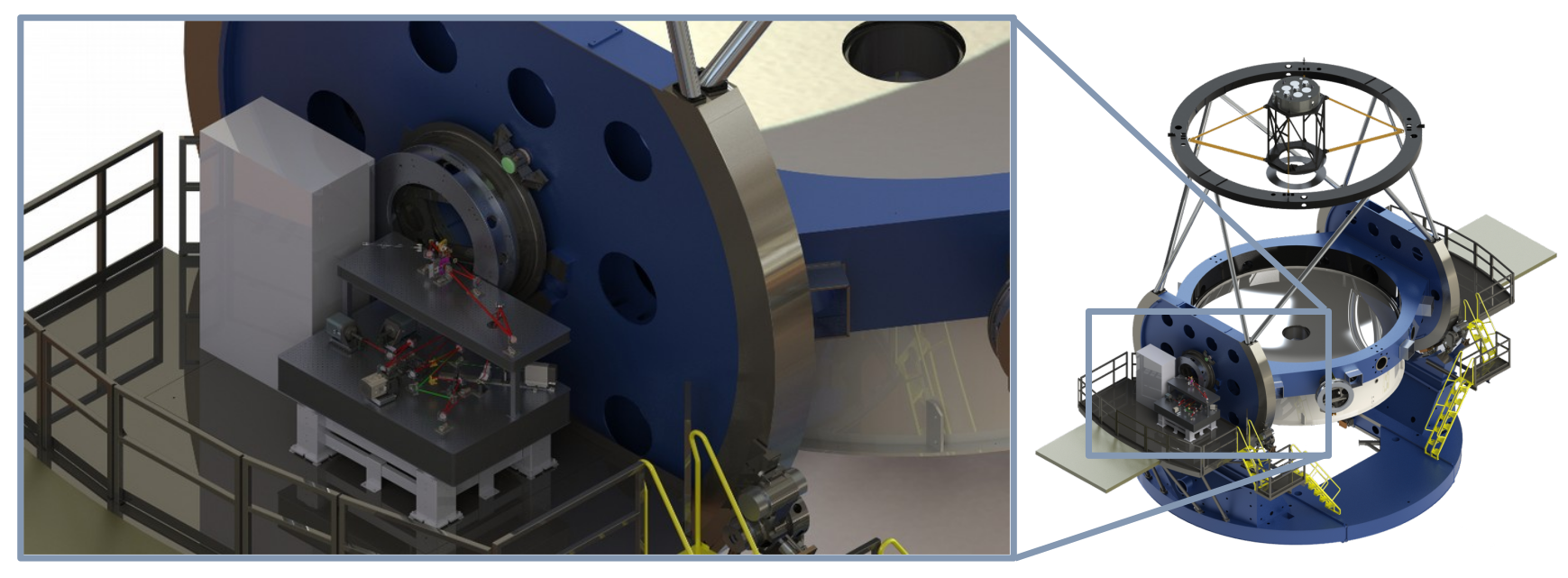}
   \caption{When observing at LCO, MagAO-X occupies the Nasmyth platform of the Magellan Clay 6.5 m telescope. \label{fig:platform_inset}}
\end{figure}

\begin{figure}[t]
   \centering
   \includegraphics[width=6in]{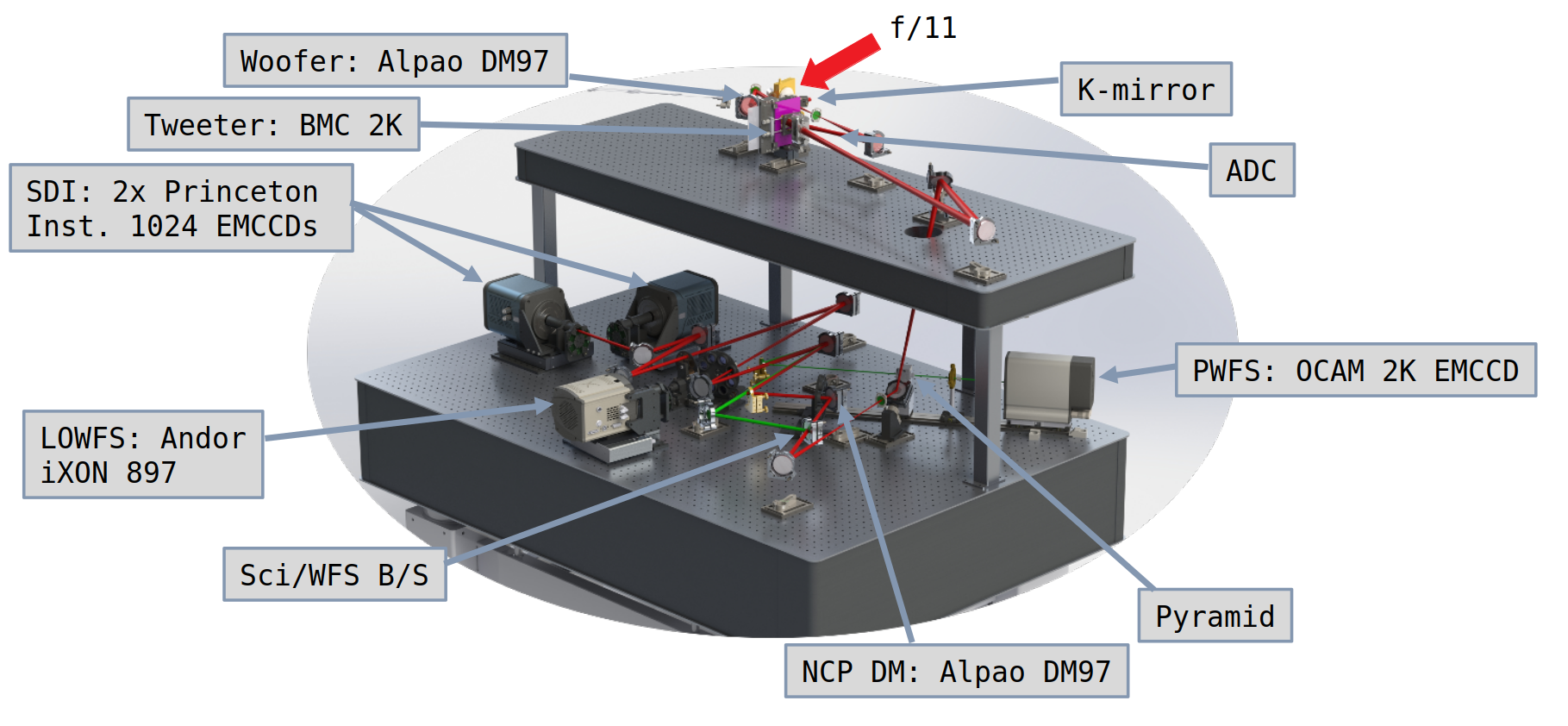}
   \caption{Overview of the MagAO-X opto-mechanical layout. \label{fig:labeled}}
\end{figure}

MagAO-X operates on the Nasmyth platform of the 6.5 Magellan Clay telescope (Figure \ref{fig:platform_inset}).  The instrument consists of a two-level optical bench (Figure \ref{fig:labeled}). It is a woofer-tweeter system, with a 97-actuator woofer and 2040-actuator tweeter.  The pupil illuminates 48 actuators across in the short direction, giving a 13.5 cm actuator pitch.  With Clay's 29\% central obscuration, roughly 1640 actuators are illuminated.  The top-level contains the k-mirror derotator and atmospheric dispersion compensator.  The bottom level  houses the pyramid wavefront sensor (PWFS) as well as a Lyot-coronagraph system feeding a dual-EMCCD simultaneous differential imaging (SDI) system.  The coronagraph contains a third deformable mirror with 97 actuators downstream of the beamsplitter, which enables non-common path offloading without offsetting the PWFS.  A visitor station is currently occupied by the VIS-X spectrograph\cite{2021SPIE11823E..06H}.  All of these components and subsystems are described in great detail in the references and documents noted above.

\section{LAB TESTBED}

MagAO-X is designed for routine shipment between LCO and the University of Arizona in Tucson.  Due to the COVID-19 Pandemic, it underwent an extended period of laboratory testing and upgrades between December 2019 and April 2022.  Notable examples of the work done during this time are the implementation of Focus Diversity Phase Retrieval (FDPR)\cite{2021JATIS...7c9001V}.  FDPR is now used at the telescope to flatten the DMs to maximize Strehl at the focal plane, and the AO is then calibrated around this point.  This procedure significantly improves on-sky Strehl.

Lab development of advanced control algorithms has also been a major emphasis of MagAO-X development.  The Data Driven Subspace Predictive Control (DDSPC) algorithm was implemented on the instrument and demonstrated in the lab.\cite{2021JATIS...7b9001H}.  We began on-sky testing in the 2022 run (Haffert et al., in prep).  In these proceedings we present new focal plane WFS techniques developed in the lab.\cite{Haffert_2022}.  MagAO-X can also be used for remote research and development, as demonstrated by the testing of a reinforcement learning algorithm.\cite{2022arXiv220507554N}

MagAO-X is also part of the High-Contrast Adaptive Optics Testbed (HCAT) for the Giant Magellan Telescope (GMT), part of the overall program to develop segment phase sensing and control strategies for the GMT\cite{Demers_2022}.  As part of HCAT MagAO-X serves as the AO system and houses phase sensing and control experiments.\cite{2020SPIE11448E..2XH,Hedglen_2022,Kautz_2022,10.1117/1.JATIS.8.2.021513,10.1117/1.JATIS.8.2.021515}

\section{ON-SKY RESULTS}
MagAO-X returned to LCO in April, 2022, following an extended laboratory upgrade and testing period (see below).  The telescope run from 09 to 23 April included 5 nights of commissioning, and 10 nights of shared-risk science verification allocated to members of the Magellan Consortium.  

\begin{figure}[h]
   \centering
   \includegraphics[width=4in]{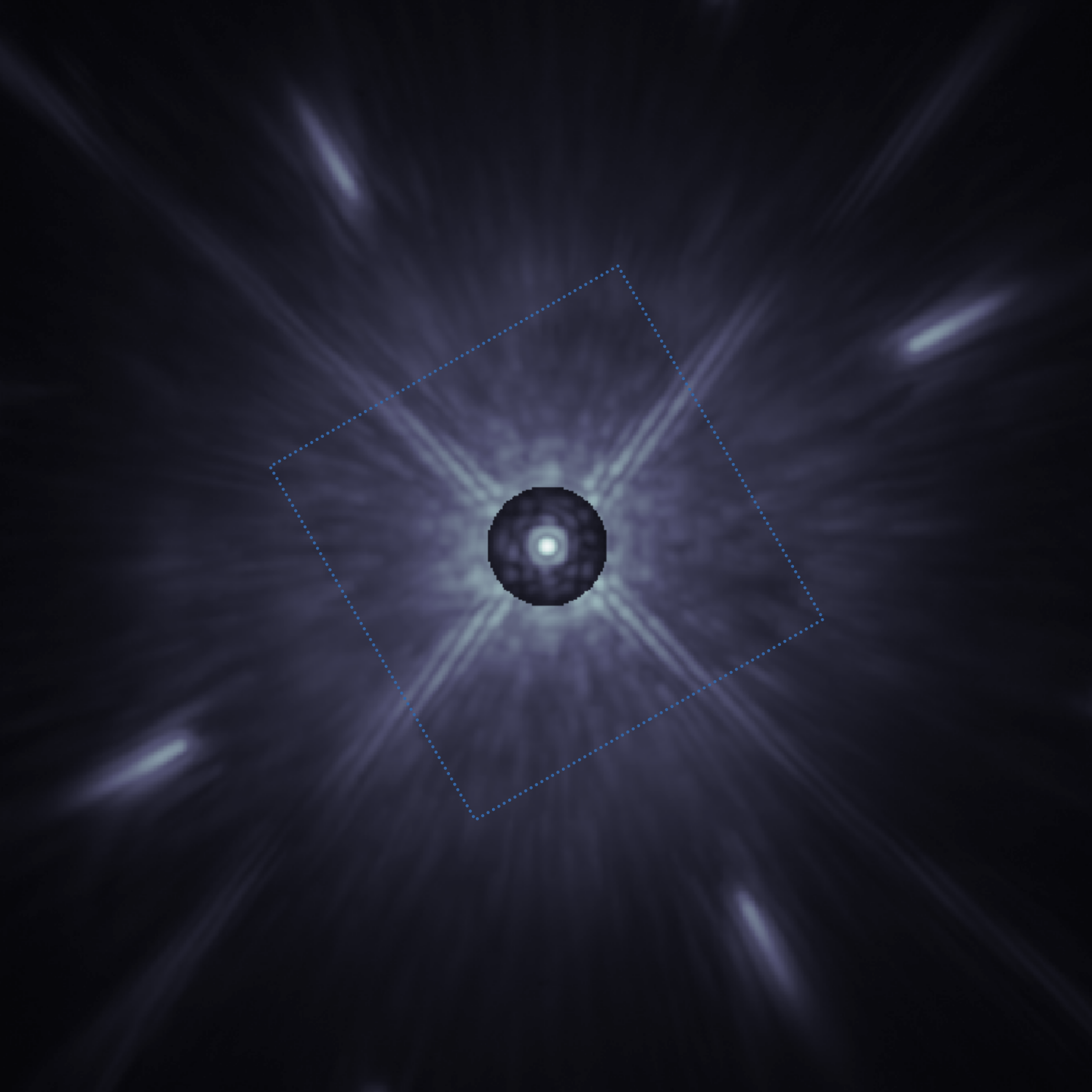}
   \caption{On-sky PSF at z' ($908$ nm), taken in 0.5'' to 0.6'' seeing on a 1st magnitude guide star.  The central circular region is on a different stretch to show detail in the core.  The blue square is 22 $\lambda/D$ in radius, highlighting the expected control region for the 1376 modal basis in use.  The control region is evident showing excellent AO performance.  The four elongated speckles between the spider diffraction arms are due to diffraction from the MEMS surface. \label{fig:z-band-dark-hole}}
\end{figure}

During this 2nd commissioning run we demonstrated significant improvements in instrument performance and operations compared to our first-light campaign.  These included procedures for flattening the instrument to maximize focal plane Strehl using our internal source each night, closed-loop pupil alignment control, the implementation of modal offloading to the woofer (instead of the actuator basis).  Together these improvements resulted in a more stable instrument compared to 2019, and enabled us to routinely close the loop on 1376 modes (compare to $\sim900$ at first light).  The result is shown in Figure \ref{fig:z-band-dark-hole}.

\begin{figure}[h]
   \centering
   \includegraphics[width=3in]{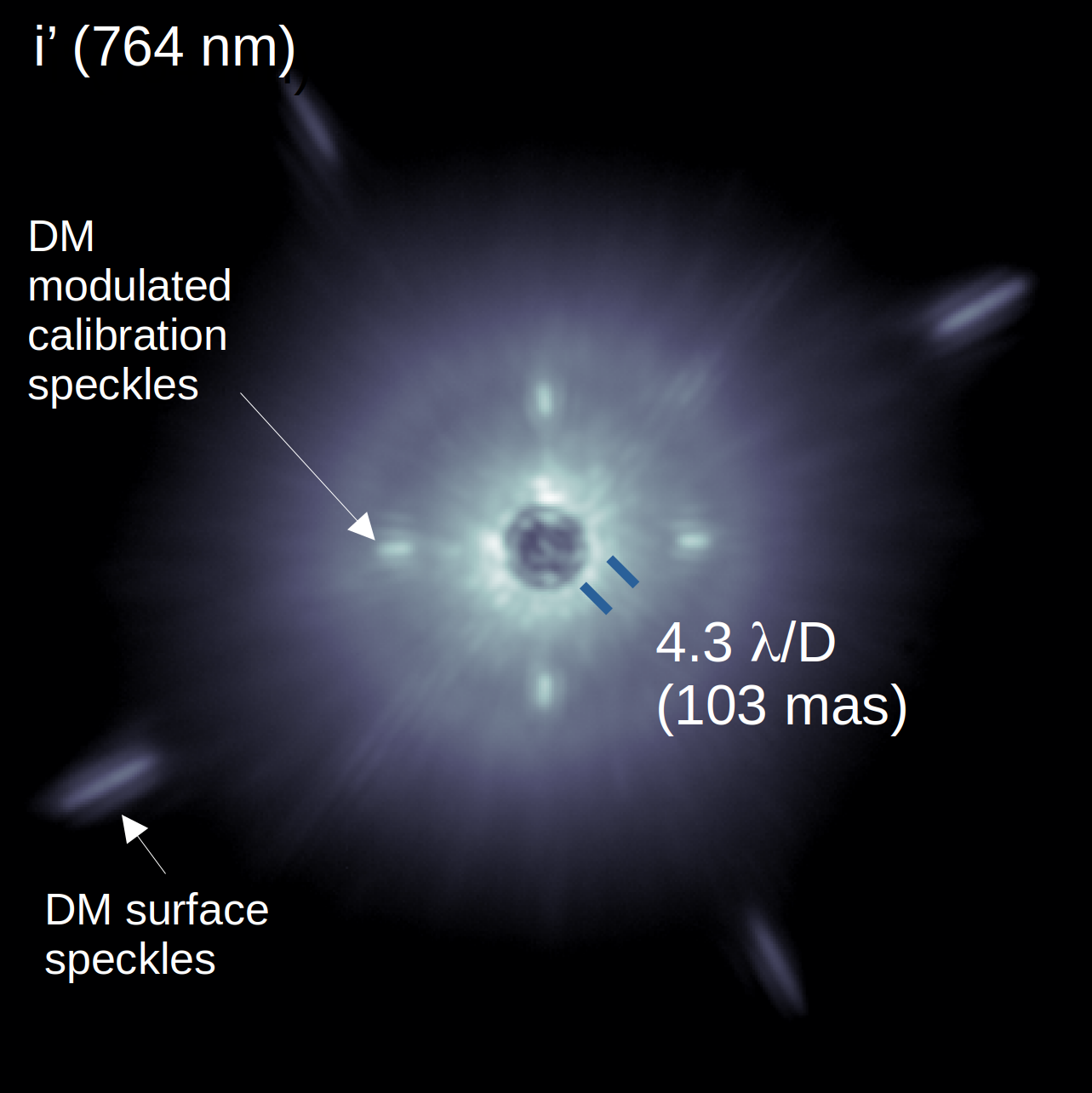}
   \caption{Lyot coronagraph observation with MagAO-X.  The central dark region is created by the 4.3 $\lambda/D$ focal plane mask.  The circular dark hole corresponds to the control region of the AO system.  Inside this dark hole, 4 speckles are visible which were created by modulating the tweeter DM.  These are used for astrometric registration and photometric calibration. \label{fig:coronagraph}}
\end{figure}

\begin{figure}[h]
   \centering
   \includegraphics[width=6.5in]{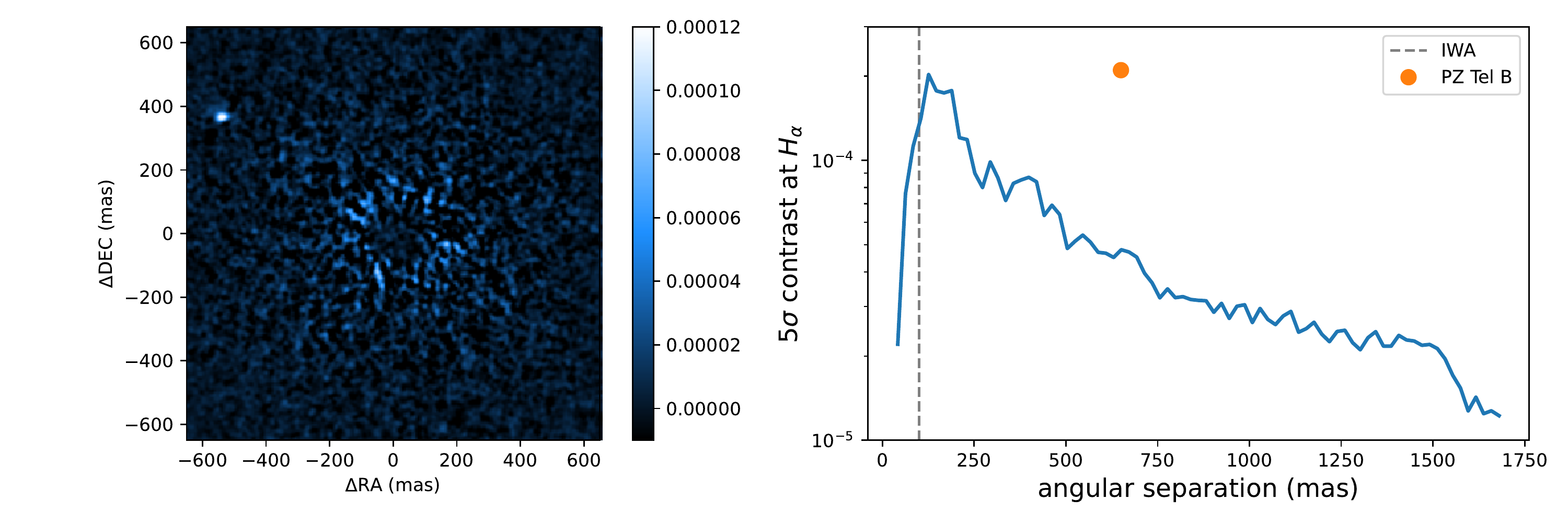}
   \caption{Coronagraphic observation of PZ Tel b at H$\alpha$.  The plot shows the calibrated contrast curve for this observation.  Haffert et al., in prep. \label{fig:pztel}}
\end{figure}

\begin{figure}[h]
   \centering
   \includegraphics[width=5in]{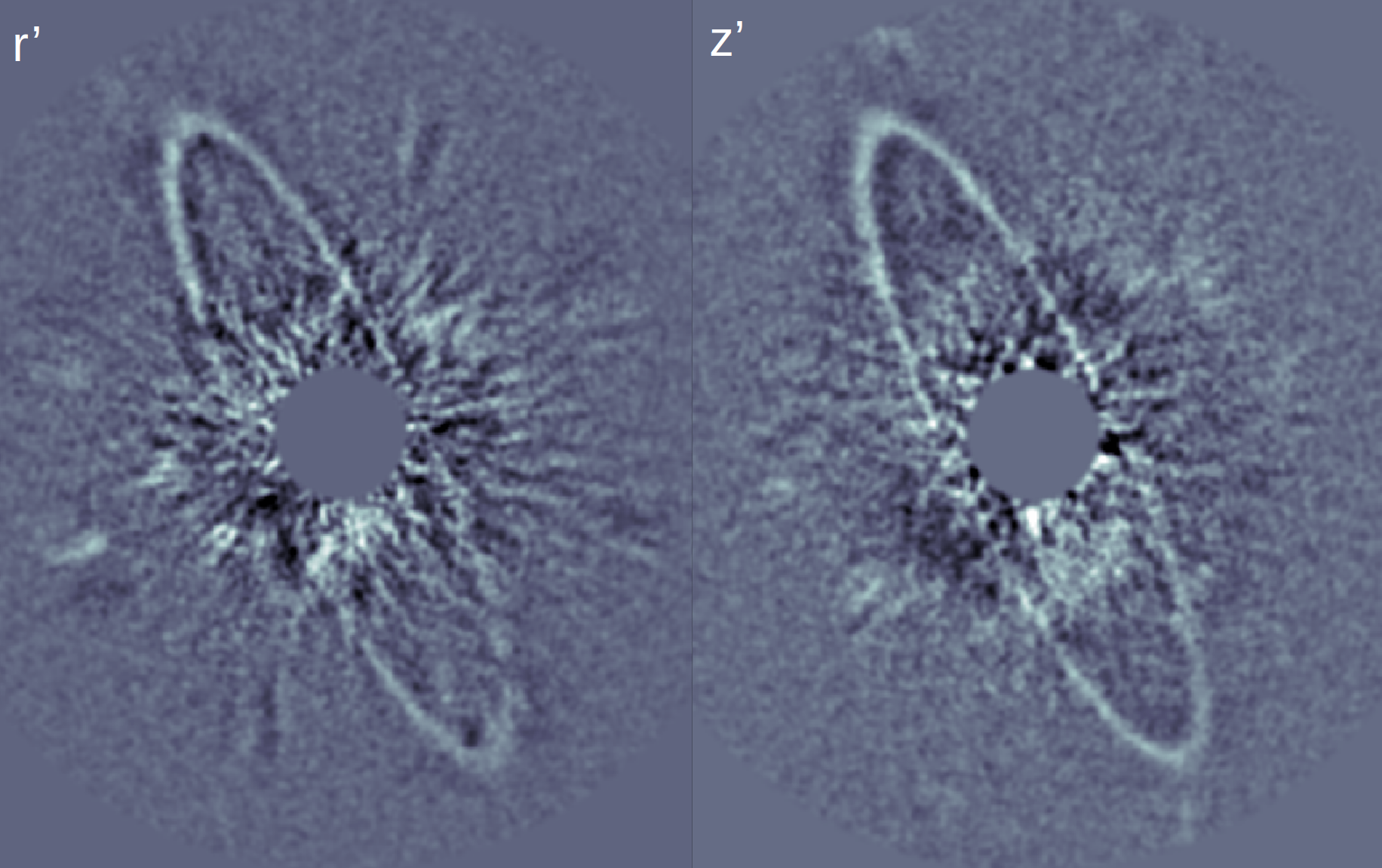}
   \caption{The well known debris disk around HR 4796 A, observed with the Lyot coronagraphs of MagAO-X.  Weinberger et al., in prep. \label{fig:hr4796A}}
\end{figure}

\begin{figure}[h]
   \centering
   \includegraphics[width=5in]{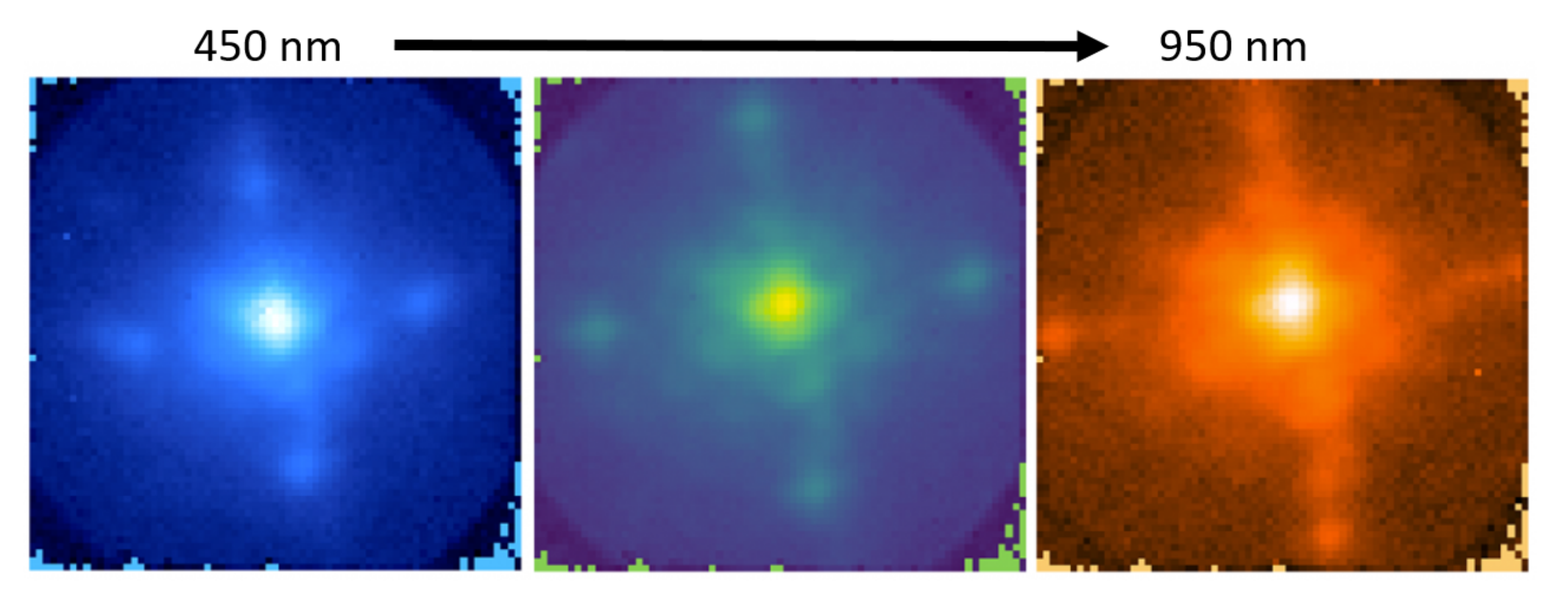}
   \caption{Commissioning of the VIS-X IFU spectrograph.  This shows the image of a star at three different wavelengths.  The four bright speckles forming a cross are DM calibration speckles.  Haffert et al., in prep. \label{fig:visx}}
\end{figure}

We also commissioned several new observing modes.  Our Lyot coronagraph system was demonstrated.  We currently support classical Lyot coronagraphs with opaque focal plane masks (details here: \url{https://magao-x.org/docs/handbook/observers/coronagraphs.html}). See Figure \ref{fig:coronagraph}, which included closed-loop feedback to the non-common path (NCP) DM inside the coronagraph using light rejected by the reflected focal plane masks and the low-order WFS camera.  The coronagraph was used extensively, for instance to observe the low-mass companion PZ Tel B in H$\alpha$ (Figure \ref{fig:pztel}) and the well-known HR 4796 A debris disk (Figure \ref{fig:hr4796A}). The VIS-X spectrograph\cite{2021SPIE11823E..06H} also began on-sky commissioning (Figure \ref{fig:visx});

The Strehl ratio of the image shown in Figure  \ref{fig:z-band-dark-hole} is $\sim60\%$, a large improvement over the $\sim46\%$ measured at First-light thanks to the better system flat and higher number of modes in use.  However, it is still below the design capability of nearly $90\%$ at \textit{z'}.  The main limitation is currently vibrations, where we sometimes experience a residual of up to 10 mas jitter.  The most likely culprit is wind shake, as similar behavior and dynamics have been observed with MagAO+VisAO.  We are investigating whether this is exacerbated by our pointing offloads to the telescope mount, which occur as often as once per second.

\begin{figure}[h]
   \centering
   \includegraphics[width=4in]{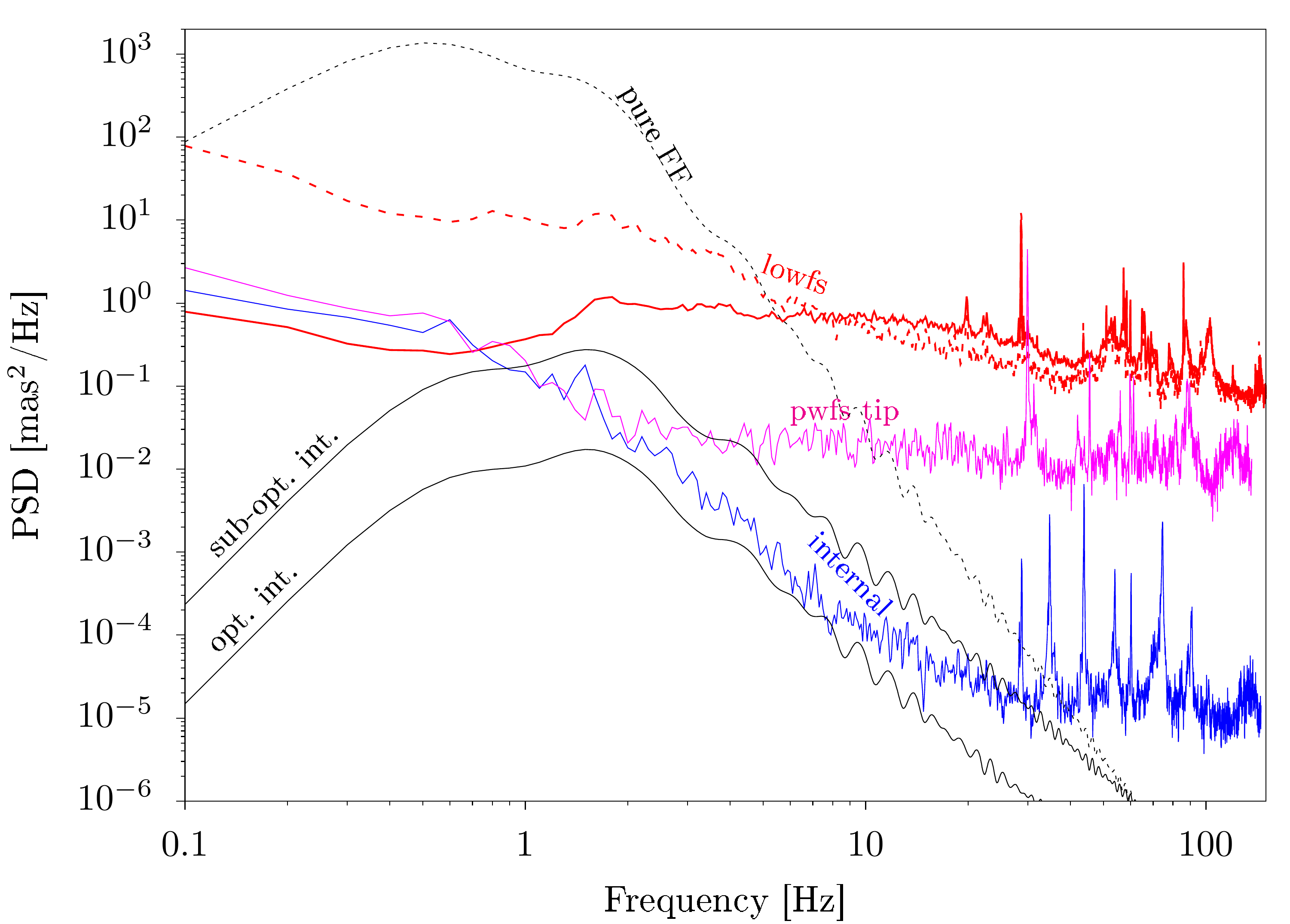}
   \caption{MagAO-X vibration power-spectra.  Black curves show expectations from frozen-flow turbulence, dashed is open-loop, solid closed-loop with optimized and sub-optimum integral control.  The blue curve shows vibrations measured on the internal source while installed at the telescope.  The pink curve is a measurement using the PWFS tip viewing camera (``camtip'').  The dashed red curve is a measurement on a different night using the LOWFS camera.  The solid red curve shows the result of closing the LOWFS loop with feedback to the NCP DM inside the coronagraph. \label{fig:vib}}
\end{figure}

\section{UPGRADES AND PHASE II}
MagAO-X was returned to Tucson for additional lab testing, HCAT testbed support, and further upgrades.  The final major item from our original plan is to test the DARKNESS MKID IFU behind MagAO-X\cite{Swimmer_2022}.  We anticipate conducting an on-sky demo with DARKNESS at LCO in early 2023.

We have also begun a significant upgrade program called ``MagAO-X Phase II''.  We plan to upgrade the real-time and instrument control computers and GPUs to reduce compute latency and enable additional real-time processing.  We are also procuring new CMOS cameras for low-order WFS, including a new Lyot-LOWFS capability.  The Lyot-LOWFS is key for the phase induced amplitude apodization complex mask coronagraph (PIAACMC) being installed as part of this upgrade.  Finally, we are upgrading the NCP DM to a 1000 actuator MEMS device to enable FPWFS\cite{Haffert_2022} without the need for PWFS offloading.

\section{CONCLUSION}

MagAO-X has now undergone two commissioning runs at LCO, and has begun supporting science observations.  Significant improvements in performance were realized between first-light in Dec, 2019 and the 2nd (COVID-delayed) commissioning run in April, 2022.  MagAO-X is capable of many exciting science cases, especially utilizing its coronagraphic modes of observation.  The instrument will continue to serve as an invaluable laboratory testbed, supporting ongoing high contrast imaging technology development as well as supporting phase sensing and control development for GMT.

\clearpage

\acknowledgments 
 
We are very grateful for support from the  NSF MRI Award \#1625441 (MagAO-X).  The Phase II upgrade program is made possible by the generous support of the Heising-Simons Foundation. 

\bibliography{report} 
\bibliographystyle{spiebib} 

\end{document}